\documentclass[
    ,final            
  ]
  {aipproc}

\layoutstyle{6x9}
\usepackage{graphicx}

\begin{document}

\title{Detecting population III galaxies\newline with HST and JWST}


\author{E. Zackrisson}{address={Department of Astronomy, Stockholm University,\\ Oscar Klein Center, AlbaNova, Stockholm SE-106 91, Sweden}}

\begin{abstract}
A small fraction of the atomic-cooling halos assembling at $z<15$ may form out of minihalos that never experienced any prior star formation, and could in principle host small galaxies of chemically unenriched stars. Since the prospects of detecting isolated population III stars appear bleak even with the upcoming {\it James Webb Space Telescope} (JWST), these population III galaxies may offer one of the best probes of population III stars in the foreseeable future. By projecting the results from population III galaxy simulations through cluster magnification maps, we predict the fluxes and surface number densities of pop III galaxy galaxies as a function of their typical star formation efficiency. We argue that a small number of lensed population III galaxies in principle could turn up at $z\approx 7$--10  in the ongoing {\it Hubble Space Telescope} survey CLASH, which covers a total of 25 low-redshift galaxy clusters. 
\end{abstract}

\maketitle


\section{Introduction}
\begin{figure}
\includegraphics[height=.4\textheight]{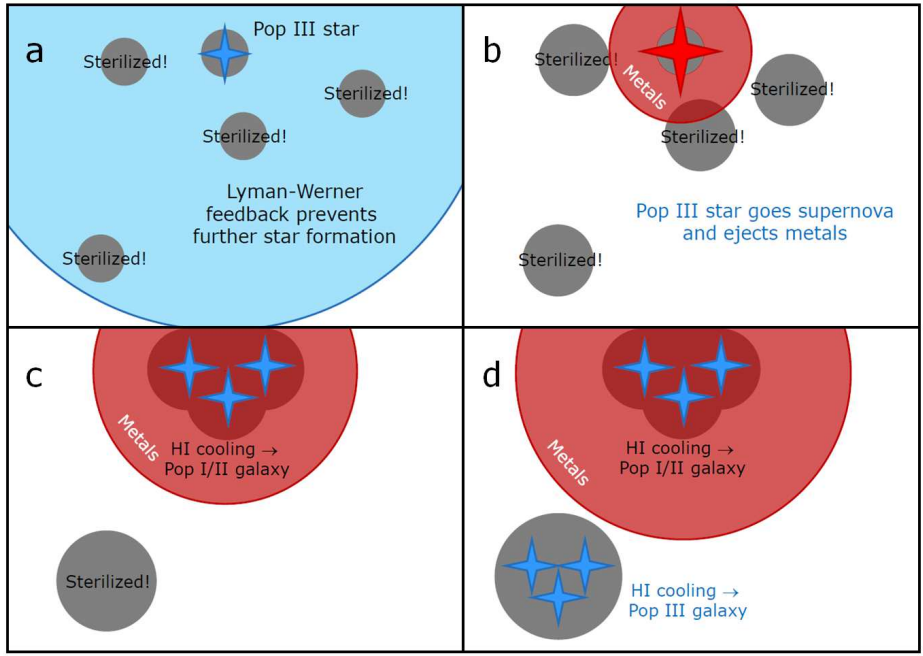}
\caption{Schematic illustration of the formation of a pop III galaxy at high redshift. {\bf a)} A minihalo forms one or several pop III stars, which emit Lyman-Werner radiation (blue region) and inhibit star formation in adjacent minihalos. {\bf b)} The pop III stars explode as supernovae and eject metals (red region) into the surrounding medium. {\bf c)} The pop III minihalo merges with a number of nearby minihalos, reaches the HI cooling limit required for prolonged star formation and forms a ``first galaxy''. Due to metal enrichment provided by the pop III star/s in one of the progenitor minhalos, the galaxy becomes a pop II/I object. {\bf d)} A rare, previously sterilized minihalo that has remained chemically pristine reaches the HI cooling limit and forms a pop III galaxy.}
\label{schematic}
\end{figure}

The first generation of population III stars (hereafter pop III) is predicted to form in isolation or in small numbers within dark matter minihalos ($\sim 10^5$--$10^6\ M_\odot$) at $z\leq 60$ \citep[e.g.][]{Trenti&Stiavelli09}. However, even with the upcoming James Webb Space Telescope (JWST), the prospects of detecting such stars appear bleak \citep{Rydberg12}, unless they attain masses $\geq 10^2$--$10^3\ M_\odot$ due to a prolonged dark star phase \citep[e.g.][]{Freese10,Zackrisson10a,Zackrisson10b,Ilie12}. Larger numbers of pop III stars could in principle form at $z\leq 15$ within HI cooling halos ($\sim 10^7$--$10^8\ M_\odot$)  that have remained chemically pristine \citep[e.g.][]{Johnson09,Stiavelli&Trenti10}, and such pop III ``galaxies'' (a.k.a. pop III star clusters) should be significantly easier to detect. Fig.~\ref{schematic} schematically illustrates how such objects might form.

\section{Detection prospects -- unlensed fields}
\begin{figure}
\includegraphics[height=.3\textheight]{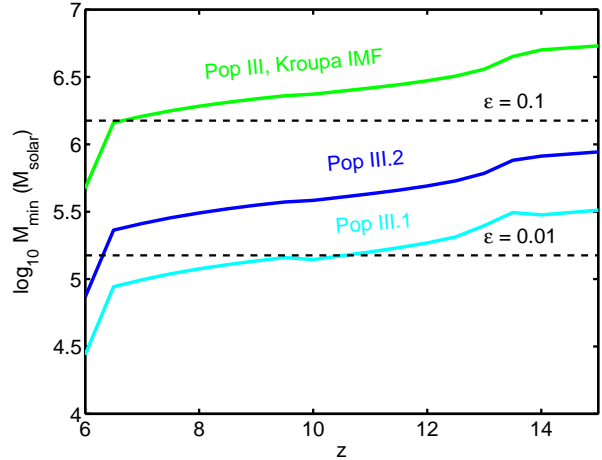}
\caption{The minimum mass in stars required for detection of pop III galaxies through JWST imaging, as a function of redshift. These limits assume 100 h JWST exposures per filter, $10\sigma$ detection in at least one JWST filter, 1 Myr old galaxies and no leakage of ionizing radiation into the intergalactic medium (type A in the notation of \citet{Zackrisson11a}). The different line colours represent different pop III stellar initial mass functions (IMFs): an extremely top-heavy IMF (cyan; characteristic mass $\sim 100\ M_\odot$); a moderately top-heavy IMF (blue; characteristic mass $\sim 10\ M_\odot$) and a stellar IMF similar to what is seen for pop II/I stars in the local Universe (green; \citet{Kroupa01}).  The dashed horizontal lines mark the total stellar masses that would correspond to star formation efficiencies of $\log_{10} \epsilon=-1$  and $\log_{10} \epsilon=-2$, assuming pop III galaxy halos of mass$10^8\ M_\odot$ and a halo baryon fraction equal to the cosmic average $f_\mathrm{b}=\Omega_\mathrm{b}/\Omega_\mathrm{M}\approx 0.15$. As seen, JWST is unlikely to detect pop III galaxies with $\log_{10} \epsilon<-2$ in unlensed fields, even in the case of an extremely top-heavy IMF.}
\label{detection_thresholds}
\end{figure}

Using the {\it Yggdrasil} spectral synthesis code \citep{Zackrisson11a} to derive the minimum star formation efficiency\footnote{here defined as the fraction of halo baryons turned into pop III stars} required to bring pop III galaxies above the JWST 10$\sigma$ photometric detection thresholds in unlensed fields, we arrive at the limits presented in Fig.~\ref{detection_thresholds} as a function of redshift (at $z\geq 7$). These results are based on very optimistic assumptions (1 Myr old galaxies, no leakage of ionizing photons into intergalactic space; 100 h JWST exposures per filter); and therefore represents hard limits on what JWST can plausibly be expected to detect. These mass detection thresholds can be converted into limits on the star formation efficiency $\epsilon$ required for a halo of a given mass. Assuming a halo mass of $10^8\ M_\odot$ regardless of redshift (also optimistic -- the typical pop III halo mass in the simulations of \citet{Trenti09} are somewhat lower than this), the star formation efficiency would need to be $\log_{10} \epsilon \geq -2$ in order to make pop III galaxies with a very top-heavy stellar initial mass function (pop III.1; characteristic mass $\sim 100\ M_\odot$) detectable. In the case of a more moderately top-heavy stellar initial mass function (pop III.2; characteristic mass $\sim 10\ M_\odot$) the limit instead becomes $\log_{10} \epsilon \geq -2$ to -1, and for a pop III stellar initial mass function similar to the pop II/I IMF in the local Universe \citep{Kroupa01}, the limit is $\log_{10} \epsilon \geq -1$. Judging from star formation efficiency ($\log_{10} \epsilon \approx -3$) predicted in the pop III galaxy simulations by \citet{Safranek12}, this is insufficient to make pop III galaxies at $z\geq 7$ detectable in unlensed fields. 

\section{Detection prospects -- lensed fields}
The use of foreground galaxy clusters as gravitational telescopes has been advocated as a possible way to detect individual pop III dark stars with JWST \citep{Zackrisson10a}, and a similar strategy can be used for pop III galaxies as well. While lensing boosts the fluxes of background objects by a factor equal to the magnification $\mu$, the background volume that is probed in such cluster lensing surveys is at the same time reduced by the same factor. This converts into a lower limit on the surface number densities for all source populations that can efficiently be detected using this method. In \citet{Zackrisson12}, we project the pop III galaxy simulation cubes of \citet{Trenti09} through the cluster magnification maps of J0717.5+3745 (J0717) -- the galaxy cluster with the largest Einstein radius so far detected \citep{Zitrin09}. This allows us to predict the lensed pop III galaxy source counts as a function of their typical star formation efficiencies. We find that cluster lensing should allow the detection of pop III galaxies with star formation efficiencies an order of magnitude lower than in unlensed fields, thereby pushing objects with $\log_{10} \epsilon \sim -3$ \citep{Safranek12} into the range detectable with JWST. If the actual pop III $\epsilon$ would be even higher ($\log_{10} \epsilon \approx -2$ to -1.5), lensed pop III galaxies could even be within reach of the ongoing HST CLASH survey \citep{Postman12}, which targets a total of 25 low-redshift galaxy clusters, including J0717. Pop III galaxy candidates at $z \geq 7$ can in principle be identified in photometric surveys  because of their unusual broadband colours \citep[e.g.][]{Pello03,Inoue11,Zackrisson11a,Zackrisson11b}. In the case of HST data, one such tell-tale colour signature can be produced by anomalously strong Ly$\alpha$ emission \citep{Zackrisson11b}. Of course, this requires that a substantial fraction of the Ly$\alpha$ photons can be transmitted through the intergalactic medium, for instance due to outflows, patchy reionization or source clustering \citep[e.g.][]{Dijkstra11,Dayal11}. 

\section{Summary and future outlook}
Depending on their typical star formation efficiencies, lensed pop III galaxies may be within reach of JWST and possibly even the HST. Such objects may, under certain circumstances be identified in multiband photometric surveys because of their unusual broadband colours. A search for such objects in the HST/CLASH survey, targeting 25 foreground galaxy clusters, is currently underway.






\bibliographystyle{aipproc}   

\bibliography{Zackrisson}

\IfFileExists{\jobname.bbl}{}
 {\typeout{}
  \typeout{******************************************}
  \typeout{** Please run "bibtex \jobname" to optain}
  \typeout{** the bibliography and then re-run LaTeX}
  \typeout{** twice to fix the references!}
  \typeout{******************************************}
  \typeout{}
 }

\end{document}